# Unveiling non-Hermitian band structures with non-Bloch supercells

Jia-Xin Zhong[*,1], Jing Lin[*,2], Kai Chen[3,4], Jing Lu[3,4,§], Kun Ding[2,‡], and Yun Jing[1,†]

[1] *Graduate Program in Acoustics, The Pennsylvania State University, University Park, PA 16802, USA*

[2] *Department of Physics, State Key Laboratory of Surface Physics, and Key Laboratory of Micro and Nano Photonic Structures (Ministry of Education), Fudan University, Shanghai 200438, China*

[3] *Key Laboratory of Modern Acoustics and Institute of Acoustics, Nanjing University, Nanjing 210093, China*

[4] *NJU-Horizon Intelligent Audio Lab, Horizon Robotics, Beijing 100094, China*

*These authors contributed equally.

§E-mail: lujing@nju.edu.cn, ‡E-mail: kunding@fudan.edu.cn, †E-mail: yqj5201@psu.edu

**Abstract**

Real-valued band structures are foundational to analyzing periodic systems within the Hermitian description and have been experimentally well-established over recent decades. In contrast, non-Hermitian systems exhibit complex band structures where both energy and momentum have imaginary parts, underpinning phenomena like the non-Hermitian skin effect and anomalous bulk-boundary correspondence that defy conventional Bloch theory. Experimentally mapping these complex bands—relating complex momentum to complex energy—and identifying their associated eigenstates is crucial for understanding these systems but remains a significant challenge. Here, we introduce a non-Bloch supercell framework designed to overcome this challenge by decoupling Bloch phase control from the imaginary part of momentum. Our method combines an exponent-flattening protocol with twisted boundary conditions, enabling system-size-independent control of imaginary momentum while preserving high-resolution Bloch phase sampling. Implemented in programmable one- and two-dimensional acoustic crystals, our approach acquires momentum-resolved complex energy surfaces and biorthogonal eigenmodes by Green's function measurements.Data obtained using this framework accurately predict open-boundary spectra and eigenstates, findings we verify through separate open-geometry experiments. Our work provides a broadly applicable experimental toolkit for exploring non-Hermitian band geometry and topology in diverse engineered classical and quantum platforms.

The foundations of quantum mechanics utilize Hermitian operators, resulting in real spectra and the conservation of probability in closed systems. Band structures, which map energy to momentum for periodic lattices based on the Bloch band theory, are a fundamental tool within this Hermitian framework. However, recognizing that most physical systems are effectively open has motivated a recent paradigm shift toward non-Hermitian descriptions. This shift is particularly prominent in the study of classical waves propagating through lattices. Non-Hermiticity leads to complex energies whose real parts are resonant frequencies and imaginary parts are decay or gain rates [1,2]. Complex spectra support unique gap and degeneracy types, known as point gaps [3–7] and exceptional points [3,8,9], underpinning phenomena [10–12] such as non-Hermitian skin effect (NHSE) [7,13–16] and sensitivity enhancement [9,17–19]. While these effects have been observed across diverse platforms [7,13,16,20–22], directly accessing complex spectra and eigenstates remains important because key features of non-Hermitian systems (e.g., van Hove singularities [23,24] and distributions of Berry curvature [25,26]) are difficult to infer reliably from emergent phenomena alone. Recent progress in measuring system-wide Green's functions (two-point correlators) has enabled direct reconstruction of complex eigenenergies and eigenstates [27].

Complex energies compel a generalization of band theory in which the full energy-momentum dispersion $E(\mathbf{k})$ and the associated eigenstates are equally essential [1,28]. Figure 1a displays one typical band structure in Hermitian settings computed under periodic boundary conditions (PBCs), accurately predicting bulk spectra and eigenstates under open boundary conditions (OBCs). By contrast, the emergence of point gaps (Fig. 1b, top) in non-Hermitian systems leads to exponential localization of bulk eigenstates at the boundary (Fig. 1b, top-right), breaking the conventional PBC-OBC correspondence [12,28]. A faithful description of the above phenomena requires complex-valued momenta $\mathbf{k} - i\boldsymbol{\mu}$ (Fig. 1b, left), yielding modes of the form $e^{\boldsymbol{\mu} \cdot \mathbf{r}}$ times Bloch waves (Fig. 1b, bottom-right)—non-Bloch eigenstates [29–35]. The non-Hermitian band structure, defined as the mapping from complex momentum to complex $E$ together with its non-Bloch eigenstates, therefore encodes all information needed to predict open-boundary behaviour, making their direct experimental access crucial.

However, standard momentum-resolved spectroscopies, such as angle-resolved photoemission spectroscopy ) [36–39], inelastic neutron scattering [40–42], and photon-induced near-field electron microscopy [43], do not provide the imaginary component of momentum $\boldsymbol{\mu}$. Because $\boldsymbol{\mu}$ cannot be retrieved by Fourier analysis alone, probing non-Hermitian band structures requires controlled implementation and readout of both real and imaginary parts of momentum, $\bm{k}$ and $\bm{\mu}$, with sufficient sampling density and operating range. As a result, complete experimental reconstruction of non-Hermitian band structures in arbitrary dimensions has remained an open challenge.

Here we introduce a non-Bloch supercell framework that enables direct experimental reconstruction of non-Hermitian band structures. We imprint $\bm{k}$ via twisted boundary conditions (TBCs) at the supercell scale, while independently tuning $\bm{\mu}$ through engineered intercell hoppings, allowing scale-free control of $\bm{\mu}$ (independent of supercell size) alongside high-resolution control of $\bm{k}$. Combined with Green's-function-based reconstruction of complex eigenenergies and eigenstates [27], this framework yields non-Hermitian band structures, encompassing the high-accuracy complex momentum-energy map and non-Bloch eigenstates. Implemented in programmable one-dimensional (1D) and two-dimensional (2D) acoustic crystals, our framework resolves Fermi points in the complex momentum context, and reveals complex-eigenenergy features such as point-gap patterns and spectral potentials. With the above complete information on non-Hermitian band structures, our framework is able to predict OBC spectra and eigenstates, which have been further verified experimentally by open-boundary systems.

To illustrate the necessity of non-Bloch supercells, we start with an $N_x \times N_y$ supercell with TBCs, where a phase factor $e^{i\theta_{x,y}}$ with $\theta_{x,y} \in (-\pi, \pi]$ is applied across each supercell boundary (Fig. 1c, top). This construction folds the primitive-cell band structure, which is usually calculated over the Brillouin zone (BZ) parameterized by $\bm{k} = (k_x, k_y)$. Increasing supercell sizes indicates larger sampling points $N_{k_{x,y}}$ in BZ along each direction, i.e., $N_{k_m} \propto N_m$ with $m = x, y$. Analytical continuation $\theta_m \to \theta_m - i\eta_m$, which is equivalent to multiplying the inter-supercell hopping along direction $m$ by $e^{\eta_m} e^{i\theta_m}$ (Fig. 1c, bottom), preserves $N_{k_m} \propto N_m$, while grants access to imaginary parts of momenta $\bm{\mu} = (\mu_x, \mu_y)$ at the primitive-cell level as $\mu_m = \eta_m/N_m$. Given $R(\eta_m)$, where $R$ denotes the operating range of a variable, $R(\mu_m)$ is inversely proportional to $N_m$, i.e., $R(\mu_m) \propto N_m^{-1}$. Consequently, $k$-resolution and $\mu$-range are interlocked: larger supercells provide finer $k$ sampling but a narrower accessible $\mu$ range, and vice versa. This $k$-$\mu$ trade-off motivates a design that decouples a wide $\mu$-range from high $k$-resolution.

To cleanly separate the roles of $\bm{k}$ and $\bm{\mu}$, we revisit non-Bloch band theory. Analytically continuing the Bloch factor $\bm{\beta} \equiv e^{ik+\mu}$ gives a $\bm{\mu}$-parameterized non-Bloch Hamiltonian $H_{\bm{\mu}}(\bm{k}) \equiv H(\bm{\beta})$, with $(\bm{k}, \bm{\mu}) \in \text{BZ} \times \mathbb{R}^d$ and $d$ being the system dimension. This yields spectra $E^\mu(\mathbf{k})$ and eigenstates $\psi_\mathbf{k}^\mu$. Relative to the PBC case ($\boldsymbol{\mu} = 0$), $\boldsymbol{\mu}$ acts as an independent deformation parameter, decoupled from $\mathbf{k}$, that reshapes both eigenvalues and eigenstates. Accordingly (Fig. 1d), we view the non-Bloch Hamiltonian as a mapping $H_{\boldsymbol{\mu}}: \mathbf{k} \to E$ from BZ to complex spectrum with attached eigenstates. For fixed $\boldsymbol{\mu}$, its image is $E^\mu(\mathbf{k})$ and the set $\Sigma = \{E^\mu(\mathbf{k}) \in \mathbb{C} \mid \boldsymbol{\mu} \in \mathbb{R}^d, k \in \text{BZ}\}$ defines the spectral manifold. Conversely, for a chosen $E_b$ (Fig. 1d, right) and given $H_{\boldsymbol{\mu}}$, the preimage on the BZ is the roots of $\det[E - H_{\boldsymbol{\mu}}(\mathbf{k})] = 0$ (red markers in Fig. 1d, left). We coin them as non-Bloch Fermi points (NBFPs), the generalization of Fermi surfaces to non-Hermitian settings, and denote

$$K_{\boldsymbol{\mu}}(E_b) = \{\mathbf{k} \in \text{BZ} \mid \det[E_b - H_{\boldsymbol{\mu}}(\mathbf{k})] = 0\}. \tag{1}$$

This treatment preserves the full information of $H(\boldsymbol{\beta})$ while explicitly disentangling $\boldsymbol{\mu}$ from $\mathbf{k}$.

With this separation, we now introduce a non-Bloch supercell scheme to resolve the $k$-$\mu$ trade-off. Concretely, we promote $\boldsymbol{\mu}$ from an implicit imaginary part of momentum to an independent, experimentally controllable deformation parameter, and encode it into real-space hoppings via a non-unitary transformation $D_{\boldsymbol{\mu}}$ [35]. For a tight-binding Hamiltonian $H = \sum_{i,j} \mathbf{t}_{i-j} c_i^\dagger c_j$ with unit-cell indices $i, j$ and hopping matrices $\mathbf{t}_{i-j}$, such a transformation $D_{\boldsymbol{\mu}}: \tilde{c}_i^\dagger \to e^{-\boldsymbol{\mu} \cdot \mathbf{R}_i} c_i^\dagger, \tilde{c}_i \to e^{\boldsymbol{\mu} \cdot \mathbf{R}_i} c_i$ yields

$$H_{\boldsymbol{\mu}} = D_{\boldsymbol{\mu}}^{-1} H D_{\boldsymbol{\mu}} = \sum_{i,j} \tilde{\mathbf{t}}_{i-j} \tilde{c}_i^\dagger \tilde{c}_j, \quad \tilde{\mathbf{t}}_{i-j} = \mathbf{t}_{i-j} e^{-\boldsymbol{\mu} \cdot (\mathbf{R}_i - \mathbf{R}_j)}, \tag{2}$$

where $\mathbf{R}_i$ and $\mathbf{R}_j$ are Bravais lattice vectors. Thus $\boldsymbol{\mu}$ imprints itself into each intercell hopping ($i \neq j$). Using this transformed lattice, we employ a supercell and impose TBCs without analytical continuation (Fig. 1e) because equation (2) encodes $\mu_m$ into the lattice itself. This maintains dense $k$-sampling $N_{k_m} \propto N_m$ while making the operating $\mu$-range independent of supercell size, i.e., $R(\mu_m) \propto N_m^0$. Hence, the $k$-$\mu$ trade-off is removed, and we designate it as the non-Bloch supercell. Compared to Fig. 1c, the supercell redistributes the boundary exponential factor $\eta_m$ uniformly over internal hoppings—a procedure we call *exponent dilution*—thereby substantially enlarging the experimentally accessible $\mu$-range (Fig. 1d).

We implement the non-Bloch supercell in an active acoustic lattice (Fig. 1f, left). The platform comprises an array of acoustic cavities coupled by programmable active elements, including microphones and loudspeakers, under real-time control (Supplementary Section I). The tight-binding analogy of such networks makes them ideal for engineering intricate hoppings and non-Hermitian terms [7,44–54]. Using a custom controller, we realize the required hopping profile for each $(\mathbf{k}, \boldsymbol{\mu})$. By point-exciting individual sites and recording the full multi-site response, we construct the Green's function $G_{ij}(E)$ and its diagonalization yields complex eigenenergies $E^\mu(\boldsymbol{k})$ and eigenstates $\psi_{\boldsymbol{k}}^\mu$ (Fig. 1f, right), enabling high-resolution experimental reconstructions of non-Bloch band structures by the proposed non-Bloch supercell framework.

We demonstrate the capability of the proposed framework from a 1D non-Hermitian lattice (Fig. 2a, top). The measured $E^\mu(k)$ for selected $\mu$ are shown by filled circles with their colour encoding $k$ (Fig. 2a, bottom). The excellent agreement with the calculated spectral manifold $\Sigma$, as shown by the gray surface, unveils our non-Bloch supercell framework in reconstructing non-Bloch band structures.

As stated above, the spectral manifold contains the necessary information to determine OBC spectra and eigenstates. To make such predictions from Fig. 2a, we analyze point gaps via eigenenergy winding numbers along a direction $\hat{n}$

$$w_{\hat{n}}(E, \boldsymbol{\mu}; \boldsymbol{k}_\perp) = \frac{1}{2\pi i} \oint \partial_{k_{\hat{n}}} \log \det[E - H_{\boldsymbol{\mu}}(k_{\hat{n}}; \boldsymbol{k}_\perp)] \, dk_{\hat{n}}, \qquad (3)$$

where $E$ is the reference energy. Equation (3) applies in arbitrary dimensions. We decompose $\boldsymbol{k}$ into $k_{\hat{n}}$ and $\boldsymbol{k}_\perp$, where $k_{\hat{n}} = \boldsymbol{k} \cdot \hat{n}$ selects the direction along which we probe point gaps, while $\boldsymbol{k}_\perp$ parametrizes the remaining $(d-1)$-dimensional section of BZ and only requires linear independence of $\hat{n}$. For the 1D system in Fig. 2a, $\hat{n}$ points along $x$ and $\boldsymbol{k}_\perp$ is absent. Since $w_{\hat{n}}$ reflect topological phase structures of eigenenergies relative to $E$, its value depends on $E$ across the complex plane. Figure 2b shows the resulting measured distributions for the 1D system. We select two $\mu$-slices from Fig. 2a and utilize the measured $E^\mu(k)$ to calculate $w(E, \mu)$, as shown by region shading colours in Fig. 2b. We refer to these $w(E, \mu)$ distributions as point-gap patterns. Self-intersections in these patterns anchor parameter pairs $(E_1, \mu = 0.1)$ and $(E_2, \mu = -0.22)$, indicating that multiple NBFPs exist at this energy (Fig. 2b, inset), which permits non-Bloch standing waves. However, self-intersection alone does not automatically belong to the OBC spectra because the OBC eigenenergy requires that it resides in point gaps for all $\boldsymbol{\mu}$, namely, $\forall \boldsymbol{\mu}, \exists (\hat{n}, \boldsymbol{k}_\perp)$ such that $w_{\hat{n}}(E, \boldsymbol{\mu}; \boldsymbol{k}_\perp) \neq 0$ (Supplementary Section II) [55].

Although our non-Bloch supercell extends the experimentally accessible $\mu$-range, access to all $\boldsymbol{\mu}$ remains constrained. We therefore employ the spectral potential [2], which can be computed alongside eigenenergy winding numbers for a measured discrete set of eigenenergies $E_\nu^\mu$ with $\nu = 1, \ldots, N_E$,

$$\phi(E, \boldsymbol{\mu}) = \int \left(\frac{dk}{2\pi}\right)^d \log|\det[E - H_{\boldsymbol{\mu}}(\boldsymbol{k})]| = \frac{1}{N_E} \sum_\nu \log|E - E_\nu^\mu|, \qquad (4)$$

which quantifies the logarithmic distance from $E$ to the spectrum. The spectral potential in the $\boldsymbol{\mu}$ space is convex, and given $E$, the presence of a plateau excludes this $E$ from OBC spectra, consistent with previous criteria from point gaps [56]. With incomplete $\mu$ coverage, Fig. 2c displays $\phi$ at $E_1$ and $E_2$ for several experimentally accessible $\mu$ values, with insets showing point-gap patterns. The absence of a plateau for $E_1$ (Fig. 2c, top) forecasts that $E_1$ is an OBC eigenenergy. In contrast, $\phi$ for $E_2$ exhibits a plateau (Fig. 2c, bottom), explaining that it is not an OBC eigenenergy.

Since spectral potential helps to pin down the OBC spectral range, together with its convex characteristic, identifying its minimum and the local behaviour around it becomes essential when judging a plateau. This motivates analyzing its $\mu$-gradient $\boldsymbol{g} = \nabla_\mu \phi(E, \boldsymbol{\mu})$, which relates to the averaged winding via $\boldsymbol{g} \cdot \hat{n} = \langle w_{\hat{n}}(E, \boldsymbol{\mu}; \boldsymbol{k}_\perp)\rangle_{\boldsymbol{k}_\perp}$, where $\langle \cdot \rangle_{\boldsymbol{k}_\perp}$ denotes averaging over $\boldsymbol{k}_\perp$ (Supplementary Section III). The V-shaped cusp for $E_1$ implies $|\boldsymbol{g}| \neq 0$ near the minimum (Fig. 2c, top), indicating that $E_1$ is always in point gaps at nearby $\mu$, as also reflected by the insets. By contrast, within the plateau region for $E_2$, the gradient vanishes and $E_2$ can lie outside the point gap (Fig. 2c, bottom). Leveraging $\boldsymbol{g}$, spectral potential provides an efficient experimental handle for predicting OBC behaviour.

Using such a recipe, the determined OBC spectra for the 1D model in Fig. 2a are shown as coloured circles in Fig. 2d, and the corresponding NBFPs are depicted in Fig. 2e, which are exactly the generalized Brillouin zone (Supplementary Section IV). The $\mu$ values of NBFPs are determined by the minimum of $\phi(E, \mu)$, while their $k$ values are then read from the corresponding measured $E^\mu(k)$. To validate, we measure system-wide Green's functions of an $L = 32$ chain, and the extracted OBC spectra (triangles, Fig. 2d) closely match the supercell prediction. As a key spectral feature, the density of states $\rho(E)$ exhibits non-Bloch van Hove singularities at energies corresponding to saddle points on generalized Brillouin zones. The bottom panel of Fig. 2d exhibits $\rho(E) = \frac{1}{L} \text{Im} \sum_v (E - E_v)^{-1}$ computed from experimental OBC eigenenergies $E_v$ for $\text{Im } E = -2.2$ Hz (orange) and $-3.9$ Hz (brown). Evident peaks near $\text{Re } E = 1033$ Hz and $1048$ Hz align with saddle points identified by our supercell method (orange triangles in Fig. 2e).

Besides spectra, the eigenstates obtained by our non-Bloch supercell framework can be examined against OBC eigenstates from Green's function measurements [27]. We select $E_1 = (1039.2 - 4.4i)$ Hz in Fig. 2b, which is also marked by the red star in the experimental OBC spectra (Fig. 2d), and plot its OBC eigenstate profile $|\psi_{E_1}|$ (Fig. 2f, triangles), which is clearly localized. For extracting its localization length, we compute the $s$-resolved Fourier-Laplace transform of $\psi_E(\mathbf{r})$

$$\tilde{\psi}_E(\mathbf{s}, \mathbf{k}) = \int d\mathbf{r}[e^{-\mathbf{s}\cdot\mathbf{r}}\psi_E(\mathbf{r})]e^{-i\mathbf{k}\cdot\mathbf{r}}, \qquad (5)$$

where $\mathbf{s} \in \mathbb{R}^d$ and $\mathbf{s} + i\mathbf{k}$ is the Laplace variable. The amplitude $|\tilde{\psi}_{E_1}(s,k)|$ (Fig. 2f, bottom) clearly shows two hotspots when $s = 0.1$, matching $\mu$ values corresponding to the spectral potential minimum from Fig. 2c. To examine further, Fig. 2f displays $e^{+0.1x}\sum_k \exp(ikx)$ (circles), where $k \in K_{\mu=0.1}(E_1)$ are two NBFPs determined in Fig. 2b. Excellent agreement in eigenstates further corroborates our supercell framework for acquiring non-Hermitian band structures.

We next apply the supercell protocol to higher dimensions using a 2D lattice (Fig. 3a). It is symmetric under $x \leftrightarrow y$ exchange, so we first investigate $\mu = \mu_x = \mu_y$ to simplify interpretations. Figure 3b displays the measured $E_\nu^\mu$ from the non-Bloch supercell for three typical $\mu$ values and for all experimentally sampled $\mathbf{k}$ points. Compared with Fig. 2b, the most striking difference is that $E_\nu^\mu$ occupies a spectral area instead of a spectral line. This is a universal spectral feature for higher-dimensional non-Hermitian systems. A salient characteristic is the nonuniform distribution of $E_\nu^\mu$ in the complex plane, which reflects the number of NBFPs mapping to each energy $E$. Hence, for a given $\boldsymbol{\mu}$, we define the NBFP density $\rho_{\boldsymbol{\mu}}(E) = \sum_\nu \delta(E - E_\nu^\mu)/N_E$, which relates to the density of states for arbitrary OBC shapes via spectral moments (Supplementary Section V). It can be computed as $\rho_{\boldsymbol{\mu}}(E) = \nabla_E^2 \phi(E, \boldsymbol{\mu})/2\pi$, where $\nabla_E^2$ are in the complex $E$ plane and $\phi(E, \boldsymbol{\mu})$ is the spectral potential in equation (4), which depends only on $E$ and $\boldsymbol{\mu}$ and is independent of $\mathbf{k}$. This quantity measures the preimage (NBFP) density in the BZ near $E$. Figure 3c presents the theoretical NBFP density corresponding to Fig. 3b, and the good agreement confirms this higher-dimensional spectral feature. Beyond density, NBFPs also have definite locations in the BZ, which we can infer from the experimental data in Fig. 3b.

We examine the experimental NBFP distribution in the BZ at $E_3$ (marked by a star in Fig. 3b), shown as circles in Fig. 3d. Because multiple NBFPs imply non-Bloch standing waves and are tied to point gaps, the eigenenergy winding numbers defined in equation (3) are projection-dependent. The choice of $\hat{n}$ sets the direction along which we diagnose point gaps and non-Bloch standing waves. Figure 3d compares two projections, $\hat{n} = \hat{y}$ (top) and $\hat{n} = (\hat{x} + \hat{y})/\sqrt{2}$ (bottom), together with their corresponding $w_{\hat{n}}(\mathbf{k}_\perp = k_x)$. In both cases, we use $k_x$ to parametrize the sweeping loop in the BZ (Fig. 3d, inset). For $\hat{n} = \hat{y}$, $w_{\hat{n}}(k_x)$ jumps by 1 whenever a $k_x$-parametrized loop crosses an NBFP (Fig. 3d, top), meaning that a point gap closes or opens in the point-gap patterns (Fig. 3e). For $\hat{n} = (\hat{x} + \hat{y})/\sqrt{2}$, $w_{\hat{n}}(k_x)$ can jump by 1 or 2, depending on how many NBFPs are traversed by the $k_x$-parametrized loops (Fig. 3d, bottom). A unit jump is the same as in the $\hat{n} = \hat{y}$ case. A jump by 2 (e.g., $-1 \rightarrow +1$) occurs when the loop intersects two NBFPs simultaneously, corresponding to spectral self-intersections at the chosen energy $E_3$ (Fig. 3f). This motivates assigning a sign to each NBFP by tracing the phase gradient of the characteristic polynomial $f(\mathbf{k}) = \det[E_b - H_{\boldsymbol{\mu}}(\mathbf{k})]$ in equation (1). By examining the phase gradient of $f(\mathbf{k})$, we determine the sign of each NBFP, displayed by filled ($+1$, clockwise) and open ($-1$, counterclockwise) circles in Fig. 3d. The jump of $w_{\hat{n}}(k_x)$ is then straightforward by examining the sum of signed NBFPs traversed by the $k_x$-parametrized loops, and vice versa. Experimentally, this capability not only predicts the potential formation of non-Bloch standing waves along the concerned direction but also offers a reliable knob for further anchoring the OBC spectral ranges.

Since $w_{\hat{n}}(\mathbf{k}_\perp)$ is experimentally accessible, we again employ the spectral potential defined in equation (4) to investigate point gaps. Figure 4a shows the theoretical $\phi(E_3, \boldsymbol{\mu})$ (surface) and its $\mu$-gradient $\mathbf{g}$ (arrows), indicating the absence of a plateau. Following the 1D discussion, $E_3$ belongs to the OBC spectra because $|\mathbf{g}| \neq 0$ near the minimizer $\boldsymbol{\mu}_{min}$ of $\phi(E_3, \boldsymbol{\mu})$. These $\boldsymbol{\mu}_{min}$ play the same role as in 1D systems, and constitute the 2D generalized Brillouin zone together with the $\mathbf{k}$ values corresponding to this $(E, \boldsymbol{\mu}_{min})$. Exploiting the convexity of $\phi(E_3, \boldsymbol{\mu})$ and the relation between $\mathbf{g}$ and the averaged eigenenergy winding numbers, we experimentally locate $\boldsymbol{\mu}_{min}$ via a quasi-Newton search that builds an approximate inverse Hessian from measured $\mathbf{g}$. Starting from the PBC scenario ($\boldsymbol{\mu} = 0$), we follow the resulting trajectory (Fig. 4a, spheres). Figure 4b displays typical NBFPs (top, circles) and $w_{\hat{y}}(k_x)$ (bottom, diamonds) for five $\boldsymbol{\mu}$. To pin down $\boldsymbol{\mu}_{min}$, we compute $\langle w_{\hat{y}}(k_x) \rangle_{k_x}$, shown as dashed lines (Fig. 4b, bottom), and minimize its distance to zero. Clearly, $\langle w_{\hat{y}}(k_x) \rangle_{k_x}$ approaches zero at $\boldsymbol{\mu}_{min} = (0.42, 0.42)$ and remains nonzero nearby, consistent with the point-gap criteria and confirming the generalized Brillouin zone condition.

Having predicted OBC behaviours from non-Bloch supercells, we next deploy a finite 2D lattice and measure system-wide Green's functions. One uniqueness in high dimensions is the system geometry, which strongly affects both OBC spectra and eigenstates. By symmetry, the measured NBFPs (Fig. 4b, right) forecast non-Bloch standing waves along the $(\hat{x} \pm \hat{y})/\sqrt{2}$ directions. We therefore fabricate a parallelogram lattice with edges aligned with these directions. Figure 4c shows the measured OBC spectra (top) and the spatial intensity summed over all right eigenstates (bottom), clearly revealing the wave localization on one edge, a signature of the NHSE.

We first analyze $E_3 = (1045.1 - 6i)$Hz, and its OBC eigenstate $|\psi_{E_3}|$ is shown in the inset of Fig. 4e. Performing the 2D $s$-resolved Fourier-Laplace transform in equation (5), we obtain the amplitude $|\tilde{\psi}_{E_3}(s = s_x = s_y, \boldsymbol{k})|$ (Fig. 4d), which exhibits two pronounced hot spots on the slice $s = 0.42$, in excellent agreement with the non-Bloch supercell prediction. The amplitude $|\tilde{\psi}_{E_3}(s = 0.42, \boldsymbol{k})|$ (Fig. 4e) shows that two of the six NBFPs are strongly excited, indicating that the open boundary selectively couples to particular standing-wave pairs. We repeat the analysis for $\psi_{E_4}$ (Fig. 4f, inset) at $E_4 = (1041.2 - 6i)$Hz and exhibit $|\tilde{\psi}_{E_4}(s_x = s_y = 0.47, \boldsymbol{k})|$ in Fig. 4f. In contrast to $\psi_{E_3}$, four of the six NBFPs are prominently excited, reflecting a distinct standing-wave composition. These measurements demonstrate that, in two dimensions, OBCs selectively excite specific NBFP combinations depending on energy and lattice symmetry, and reveal that the NBFP locations in the BZ are as important as $(E, \boldsymbol{\mu}_{min})$.

In summary, we have introduced and experimentally demonstrated a scalable protocol for mapping non-Hermitian band structures. By treating Bloch phase and the imaginary part of crystal momentum as independent knobs—implemented via non-Bloch supercells with exponent-flattening—and by extracting complex eigenenergies and biorthogonal eigenmodes through the Green's function measurements, we have achieved high-resolution, momentum-resolved mapping of complex-energy surfaces and their diagnostics, including non-Bloch Fermi points, point-gap patterns, and spectral potentials. Implemented in programmable 1D and 2D acoustic crystals, the non-Bloch supercell scheme experimentally predicts OBC spectra and mode compositions, providing a compact and transferable toolkit for characterizing the geometric, topological, and spectral content of non-Hermitian bands.

Since momentum-resolved complex eigenenergies and biorthogonal eigenmodes are obtainable within our non-Bloch supercell framework, one can experimentally probe non-Hermitian quantum geometry (Berry curvature and quantum metric), track exceptional degeneracies in multiband settings, and analyze intrinsically higher-dimensional complex spectral manifolds and their distinctive geometric signatures. Our framework, which recapitulates previously synthesized artificial-boundary protocols in the single-cell limit [45,46,57], is implementation-agnostic and readily transferable to mechanical, circuit, microwave, and photonic platforms, offering a broadly applicable experimental toolbox for non-Hermitian band engineering.

**Acknowledgment**

Y. J. thanks the support of startup funds from Penn State University and NSF awards 2039463 and 195122. This work is supported by the National Key R&D Program of China (2022YFA1404500, 2022YFA1404701), the National Natural Science Foundation of China (12174072, 2021hwyq05, 12347144), and the Shanghai Science and Technology Innovation Action Plan (No. 24Z510205936).

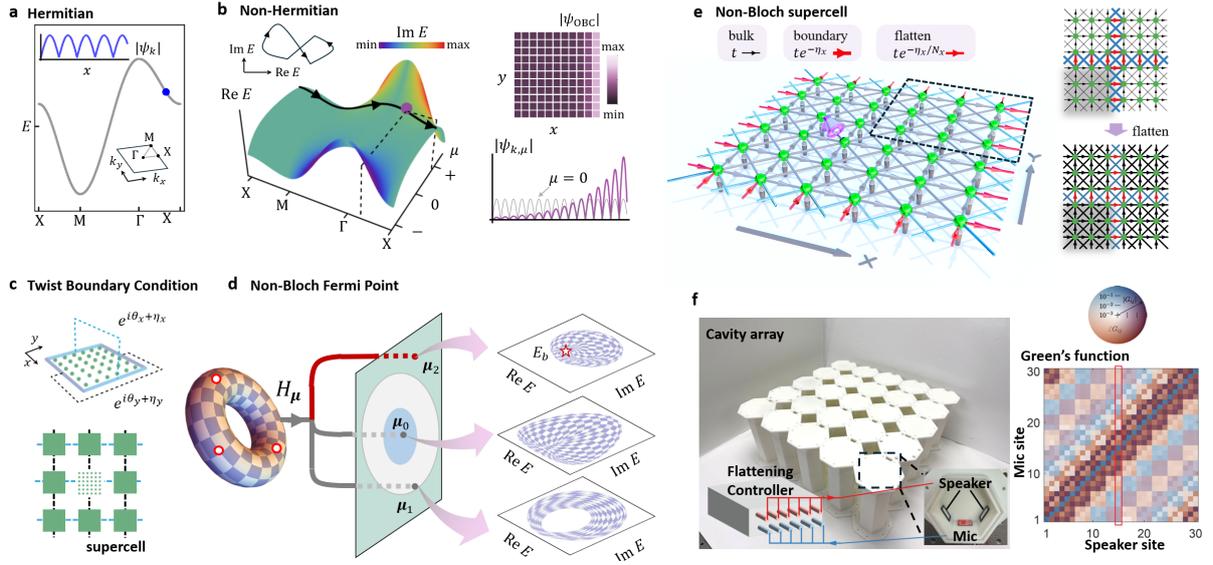

**Fig. 1 | Non-Bloch supercell framework for reconstructing non-Hermitian band structures in experiment. a**, Representative Hermitian band structure. Only one high-symmetry line of a square lattice is shown, and the inset depicts a typical Bloch eigenstate. **b**, Representative non-Hermitian band structure. The surface (left) displays real parts of eigenenergy along the high-symmetry line under analytical continuation of $\mathbf{k} \to \mathbf{k} - i\boldsymbol{\mu}$, while its colour encodes imaginary parts of eigenenergy. The right panel displays one OBC eigenstate (top) and two representative non-Bloch eigenstates (bottom). **c**, Schematic of a supercell with TBCs. The top panel shows analytical continuation $\boldsymbol{\theta} - i\boldsymbol{\eta}$ of the TBCs, and the bottom panel sketches its real-space correspondence. The number of unit cells along direction $x(y)$ is $N_{x(y)}$. **d**, Mapping $H_{\boldsymbol{\mu}}: \mathbf{k} \to E$ from BZ to complex spectrum for arbitrary $\boldsymbol{\mu}$. The red circles denote three representative values of $\boldsymbol{\mu}$, and the resulting energy spectra are visualized on the right. The preimages of a given $E_b$ for $\boldsymbol{\mu}_2$ are shown by red stars in the BZ and referred to as NBFPs. **e**, The non-Bloch supercell framework to probe non-Hermitian band structures. The top-right panel displays analytical continuation $\boldsymbol{\theta} - i\boldsymbol{\eta}$ of supercells. Red and blue thick lines highlight the effective hopping modifications induced by $\boldsymbol{\eta}$, for example, the right-forward hopping is changed as $t \to te^{-\eta_x}$. The bottom-right panel displays the non-Bloch supercell with $e^{i\boldsymbol{\theta}}$ applied at the boundaries (red and blue thin lines) and diluting $e^{\boldsymbol{\eta}}$ that alter the boundary hoppings across the entire lattice at each intercell hopping (gray lines), for example, the right-forward hopping $t \to te^{-\eta_x/N_x} = te^{-\mu_x}$. **f**, Samples and Green's function measurements. The left panel shows the cavity array, where each site hosts two speakers and a microphone for actuation and sensing. All onsite potentials and hopping terms, including $\boldsymbol{\theta}$ and $\boldsymbol{\mu}$, are dynamically programmed via a real-time controller that processes the microphone signals and drives the speakers accordingly. The right panel depicts the representative Green's function

measurement, rendering complex spectra and eigenstates observable experimentally. The horizontal and vertical axes stand for speaker and microphone sites, respectively.

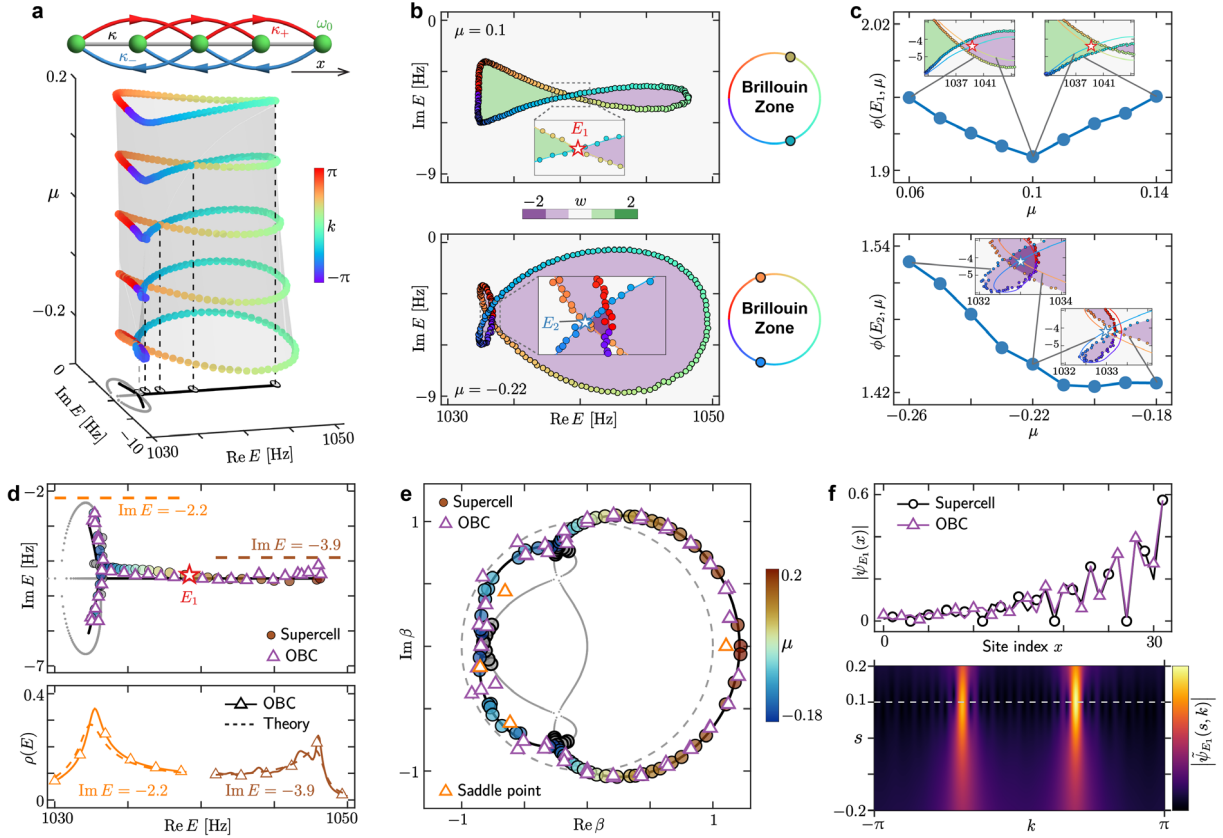

**Fig. 2 | Experimentally determined 1D non-Hermitian band structures and OBC verifications. a**, Non-Hermitian band structures with both complex-valued energies and wavenumbers. The used lattice model is $H_{1D} = \omega_0 + \kappa(\beta + \beta^{-1}) + \kappa_+\beta^{-2} + \kappa_-\beta^2$, as sketched in the top. The gray surface is from theory, while the coloured spheres display experimental results with $k$ encoded by colour. The bottom plane shows the projection of five self-intersection points, and the black line indicates the theoretical OBC spectra. **b**, Measured point-gap patterns $w(E,\mu)$ for $\mu = 0.1$ (top) and $\mu = -0.22$ (bottom). The marker colours represent $k$, while region shading encodes the value of $w$ for $E$ within each region. Insets highlight point-gap patterns near spectral self-intersection points $E_1 = (1039.2 - 4.4i)$Hz and $E_2 = (1033.0 - 4.2i)$Hz. Their corresponding $k$ preimages in the BZ, NBFPs, are shown by filled circles in the right panels. **c**, Measured spectral potential $\phi(E,\mu)$ for $E_1$ (top) and $E_2$ (bottom). A V-shaped cusp in $\phi$ indicates $E$ is an OBC eigenenergy (top); otherwise, $E$ is excluded from the OBC spectrum (bottom). Insets display evolutions of point-gap patterns with $\mu$. **d,e**, OBC spectra (**d**, top) and generalized Brillouin zones (**e**) determined from non-Bloch supercells (circles) and measured in a finite OBC lattice (triangles). The bottom panel in **d** shows the density of states, $\rho(E)$, versus Re $E$ at Im $E = -2.2$ Hz (orange) and Im $E = -3.9$ Hz (brown). The density of state features arise from saddle points highlighted in **e**. Black and gray dots represent theoretical predictions. **f**, Measured real-space wavefunction $\psi_{E_1}(x)$

(triangles, top) and $s$-resolved momentum-space wavefunction $\tilde{\psi}_{E_1}(s, k)$ (bottom) from open boundary lattices. Hotspots of $\tilde{\psi}_{E_1}(s, k)$ at $s = 0.1$ agree with NBFPs determined in **b**. Filled circles on the top depict wavefunctions determined from the non-Bloch supercell measurement. The parameters used are $\omega_0 = (1038 - 4i)$Hz, $\kappa = 4$Hz, $\kappa_+ = 2$Hz, and $\kappa_- = 0.4$Hz. The supercell is composed of $N_x = 25$ unit cells, while the size of the OBC system is $L = 32$.

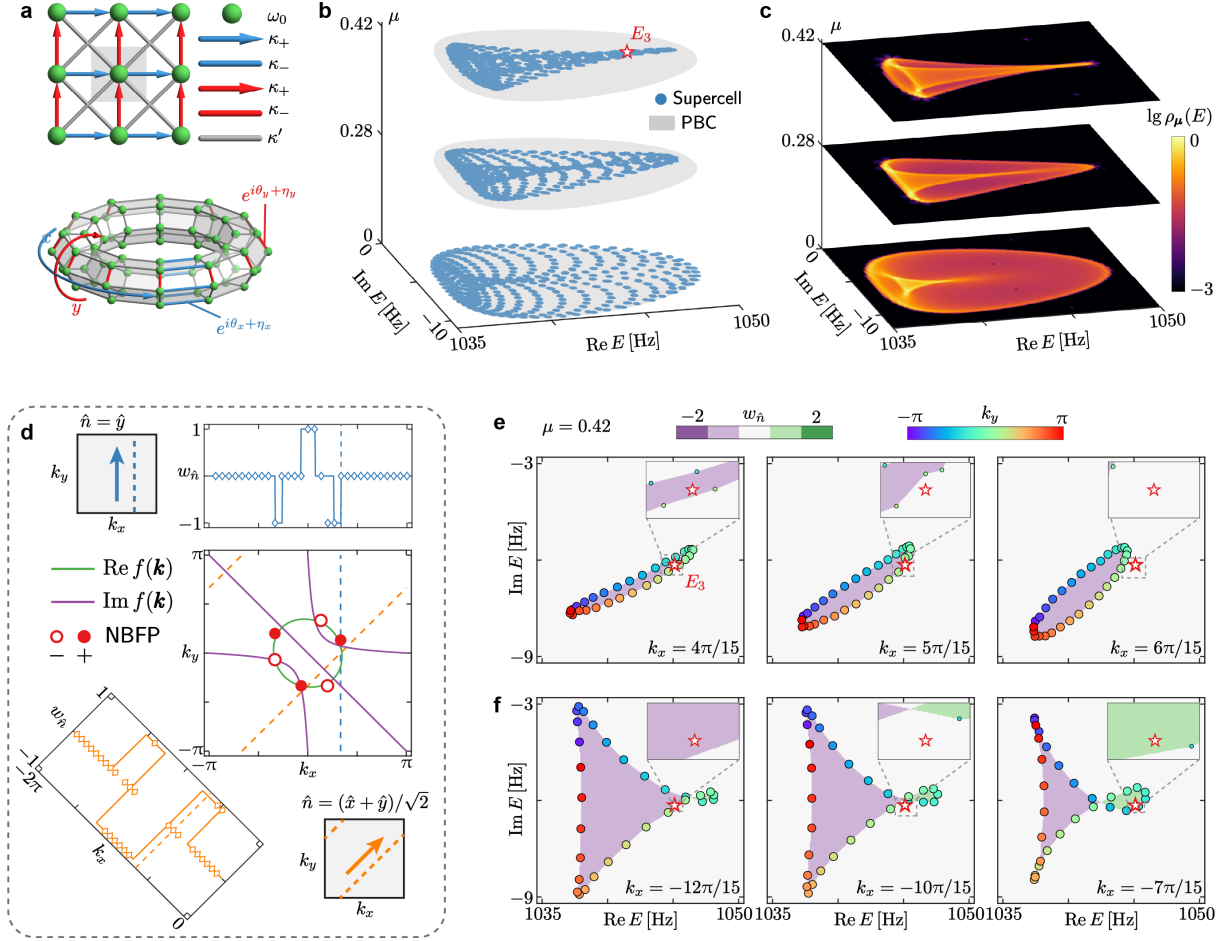

**Fig. 3 | 2D non-Hermitian band structures with projection-dependent diagnostics induced by NBFPs. a**, Schematic of a 2D nonreciprocal lattice (top) and its non-Bloch supercell (bottom). The used Hamiltonian is $H_{2D} = \omega_0 + \kappa_+(\beta_x^{-1} + \beta_y^{-1}) + \kappa_-(\beta_x + \beta_y) + \kappa'(\beta_x\beta_y + \beta_x^{-1}\beta_y + \beta_x\beta_y^{-1} + \beta_x^{-1}\beta_y^{-1})$. **b**, Measured complex-energy spectra (blue dots) for $\mu_x = \mu_y = \mu$. The gray region indicates the experimentally determined PBC spectral range. **c**, Calculated density of NBFPs, $\rho_\mu(E)$, for the corresponding $\mu$ in **b**. **d**, Measured NBFP distribution for $\mu = 0.42$ and $E_3 = (1045.1 - 6i)$Hz (red star in **b**). Six red markers denote NBFPs extracted from the experiment in **b** with open (filled) indicating their sign. Green (purple) lines show the theoretical loci $\text{Re}f(\bm{k}) = 0$ [$\text{Im}f(\bm{k})$]; their intersections are the theoretical NBFPs. The top and bottom panels display two projection directions $\hat{n}$ and the ensuing eigenenergy winding number distributions $w_{\hat{n}}(k_x)$, where $\hat{n}$ specifies the direction along which the winding number integral is performed. For example, in the top panel, the eigenenergy winding number $w_{\hat{y}}(k_x)$ is computed by integrating over $k_y$ and has distributions in $k_x$. **e**, **f**, Measured point-gap patterns $w_{\hat{n}}(E_3, \mu = 0.42; k_x)$ for $\hat{n} = \hat{y}$ (**e**) and $\hat{n} = (\hat{x} + \hat{y})/\sqrt{2}$ (**f**). The three columns show patterns as $k_x$ crosses one NBFP (**e**) or two NBFPs (**f**). Marker colour encodes $k_{\hat{n}}$ (= $\bm{k} \cdot \hat{n}$),

and region shading encodes $w_{\hat{n}}(E_3, \mu; k_x)$. The evolution in **e** displays a typical topological transition of point gaps, while the evolution in **f** corresponds to the self-intersection scenario, which behaves like Fig. 2b. The parameters used are $\omega_0 = (1040 - 6i)$Hz, $\kappa' = 0.64$Hz, $\kappa_+ = 2.72$Hz, $\kappa_- = 0.48$Hz, $N_x = 12$, and $N_y = 6$.

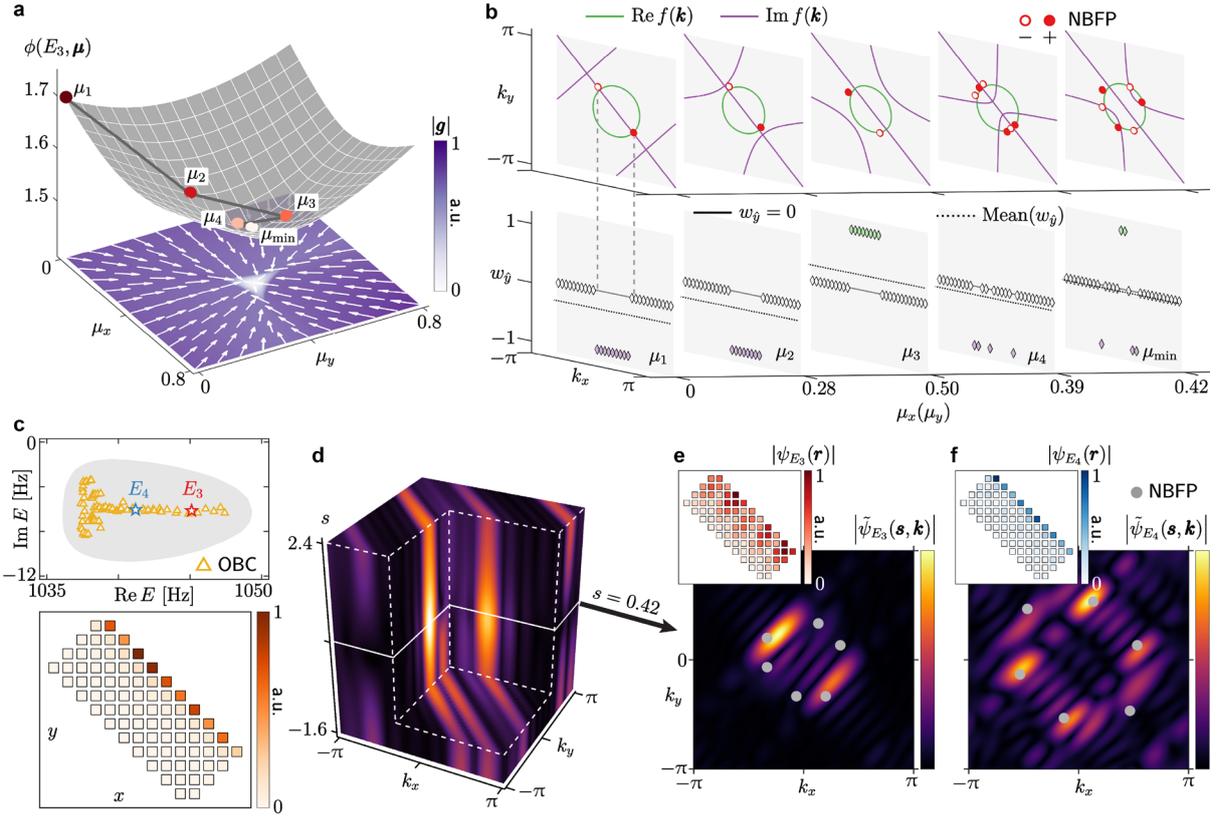

**Fig. 4 | Experimental determination of spectral-potential minima, ensuing NBFPs, and OBC validation. a**, Spectral potential $\phi(E,\boldsymbol{\mu})$ for $E_3 = (1045.1 - 6i)$Hz (red star in Fig. 3**b**). The colour map and arrows on the bottom plane show $|\boldsymbol{g}|$ and $\boldsymbol{g}/|\boldsymbol{g}|$, respectively, where $\boldsymbol{g} = \nabla_{\boldsymbol{\mu}}\phi(E,\boldsymbol{\mu})$ relates to eigenenergy winding numbers as $\langle w_{\hat{n}}(E,\boldsymbol{\mu};k_x)\rangle_{k_x} = \boldsymbol{g}\cdot\hat{n}$. The $\boldsymbol{\mu}$ value at which $|\boldsymbol{g}|$ vanishes, $\boldsymbol{\mu}_{\min}$, is marked by white spheres and coincides with the minimum of $\phi(E_3,\boldsymbol{\mu})$. **b**, Evolution of measured NBFPs (open and filled circles, top) and their eigenenergy winding numbers $w_{\hat{y}}(E_3,\boldsymbol{\mu};k_x)$ (diamonds, bottom) for the $\boldsymbol{\mu}$ values indicated in **a**. The $\boldsymbol{\mu}$ positions follow a BFGS minimizer to approach $\boldsymbol{\mu}_{\min}$. Green (purple) curves in the top row are the theoretical loci $\text{Re}f(\boldsymbol{k}) = 0$ [$\text{Im}f(\boldsymbol{k}) = 0$]. In the bottom row, the solid (dashed) horizontal line denotes $w_{\hat{y}} = 0$ [$\langle w_{\hat{y}}(E_3,\boldsymbol{\mu};k_x)\rangle_{k_x}$]. **c**, OBC spectra (top) and the spectral sum of right-eigenstate amplitudes (bottom) measured on a finite OBC lattice with 80 sites. Red and blue stars mark two selected energies, $E_3 = (1045.1 - 6i)$Hz and $E_4 = (1041.2 - 6i)$Hz. Their $\boldsymbol{\mu}_{\min}$ values, determined via non-Bloch supercells, are 0.42 and 0.47, respectively. **d**, $s$-resolved momentum-space wavefunction $|\tilde{\psi}_E(\boldsymbol{s},\boldsymbol{k})|$ for $E = E_3$ as a function of $\boldsymbol{k}$ and $s_x = s_y = s$. White solid lines mark the $s = 0.42$ slice hosting the hotspot of $|\tilde{\psi}_{E_3}(\boldsymbol{s},\boldsymbol{k})|$. **e,f**, $s$-resolved momentum-space wavefunction $|\tilde{\psi}_E(\boldsymbol{s},\boldsymbol{k})|$ for $E = E_3$ (**e**) and $E = E_4$ (**f**), evaluated at the hotspot of $|\tilde{\psi}_E(\boldsymbol{s},\boldsymbol{k})|$ for each energy. Gray circles mark NBFP positions measured by

non-Bloch supercells, and the insets display the corresponding real-space right eigenstates measured by OBC lattices.